\documentclass[prl,twocolumn,showpacs,preprintnumbers,amsmath,amssymb, superscriptaddress]{revtex4}

\usepackage{graphicx}
\usepackage{epstopdf}
\usepackage{dcolumn}
\usepackage{bm}
\usepackage{epsfig}
\usepackage{amsmath}
\begin{document}


\title{Single 3$d$ transition metal atoms on multi-layer graphene systems: electronic configurations, bonding mechanisms and role of the substrate}

\author{V. Sessi}
\affiliation{European Synchrotron Radiation Facility, 6 rue Jules Horowitz, BP 220 38043 Grenoble Cedex 9, France}
\affiliation{Institute of Semiconductor and Solid State Physics, Johannes Kepler University, Altenberger Strasse 69, A-4040 Linz, Austria}
\author{S. Stepanow}
\affiliation{Max-Planck-Institut f\"ur Festk\"orperforschung,
Heisenbergstrasse 1, 70569 Stuttgart, Germany}
\affiliation{Department of Materials, ETH Z\"urich, H\"onggerbergring 64, 8093 Z\"urich, Switzerland}
\author{A. N. Rudenko}
\affiliation{Radboud University Nijmegen, Institute for Molecules and Materials, Heijendaalseweg 135, NL-6525AJ Nijmegen, The Netherlands}
\author{S. Krotzky}
\affiliation{Max-Planck-Institut f\"ur Festk\"orperforschung,
Heisenbergstrasse 1, 70569 Stuttgart, Germany}
\author{K. Kern}
\affiliation{Max-Planck-Institut f\"ur Festk\"orperforschung,
Heisenbergstrasse 1, 70569 Stuttgart, Germany}
\author{F. Hiebel}
\affiliation{Institut N\'eel, CNRS-UJF, BP 166, F-38042 Grenoble, France}
\author{P. Mallet}
\affiliation{Institut N\'eel, CNRS-UJF, BP 166, F-38042 Grenoble, France}
\author{J.-Y. Veuillen}
\affiliation{Institut N\'eel, CNRS-UJF, BP 166, F-38042 Grenoble, France}
\author{O. \v{S}ipr}
\affiliation{Institute of Physics of the ASCR v. v. i., Na Slocance 2, CZ-182 21 Prague, Czech Republic}
\author{J. Honolka}
\email{honolka@fzu.cz}
\affiliation{Max-Planck-Institut f\"ur Festk\"orperforschung,
Heisenbergstrasse 1, 70569 Stuttgart, Germany}
\affiliation{Institute of Physics of the ASCR v. v. i., Na Slocance 2, CZ-182 21 Prague, Czech Republic}
\author{N. B. Brookes}
\affiliation{European Synchrotron Radiation Facility, 6 rue Jules Horowitz, BP 220 38043 Grenoble Cedex 9, France}

\date{\today}

\begin{abstract}
The electronic configurations of Fe, Co, Ni, and Cu adatoms on graphene and graphite have been studied by x-ray magnetic circular dichroism and charge transfer multiplet theory. A delicate interplay between long-range interactions and local chemical bonding is found to influence the adatom equilibrium distance and magnetic moment. The results for Fe and Co are consistent with purely physisorbed species having, however, different 3$d$-shell occupancies on graphene and graphite ($d^{n+1}$ and $d^{n}$, respectively). On the other hand, for the late 3$d$ metals Ni and Cu a trend towards chemisorption is found, which strongly quenches the magnetic moment on both substrates. \end{abstract}

\maketitle
%
Graphene is recognized as an excellent candidate for spintronics applications. The incorporation of magnetic elements in graphene is expected to result in hybrid systems with a variety of magnetic functionalities, given the large spin diffusion length and high carrier mobility in graphene~\cite{spin1,mobility}. Interfaces with ferromagnetic layers for example are promising for both spin injection~\cite{spin2,spin injection} and magnetic tunnel junctions~\cite{magnetoresistance} whereas single magnetic impurities attached to graphene are predicted to spin polarize graphene atoms, leading to gate tunable, indirect magnetic coupling~\cite{Castro-Neto,Sengupta}. Transition metal (TM) atoms of the fourth period provide a convenient model system to study the interactions with graphene at the atomic scale, since their outmost 3$d$-shells, accessible to spectroscopy studies, are very sensitive to the chemical and magnetic environment of the adatom.


A few scanning tunneling microscopy (STM) experiments have addressed the adsorption behavior of TM adatoms on several surface supported graphene systems: Fe on graphene/Ru(0001)~\cite{Hamburg1}; Co on graphene/SiO$_2$~\cite{Crommie}; Co, Fe and Ni on graphene on the Si-face of SiC~\cite{Elbo,graphene,Hamburg2}; Co on HOPG~\cite{Co_HOPG} and Co on graphene/Pt(111)~\cite{Brune}. First studies on the magnetic properties of single TM adatoms on supported graphene were presented only very recently~\cite{Elbo,Brune}.
From these investigations it appears that the adsorption geometries, interaction strengths and magnetic susceptibilities depend on the specific properties of the employed graphene sample system.

Standard functionals in Density Functional Theory (DFT) poorly capture dispersive interactions~\cite{a} and shortcomings are especially prominent when modeling graphene situated on underlying materials \cite{b}, which participate in long-range interactions and can greatly influence the bonding character.
However, DFT-based approaches have been extensively used to predict magnetic moments and adsorption geometry for the ideal case of TMs on free-standing graphene~\cite{Xiao, Duffy-Blackman, Johll, Krasheninnikov, Wehlig}. Within DFT the free-atom electronic configurations are never preserved, as the 4$s$ states become unfavorable in the presence of graphene and promote larger 3$d$-shell occupations. As a consequence, for the late 3$d$ elements Fe, Co and Ni, lower spin values compared to the free-atom case are expected.
In disagreement to the DFT results, a recent quantum chemical study ~\cite{Rudenko} has suggested that Co atoms might be physisorbed on graphene, with distances to the graphene layer large enough to maintain the free-atom electronic configuration $d^7$ with spin 3/2. First experimental results for Fe, Co and Ni in Ref.~\cite{Elbo} seem to hint enhanced $d$-shell occupancies, however, a rigorous quantitative
determination is missing.
\begin{figure}
\begin{center}
\includegraphics[trim=0 0 0 0,clip,width=8.5cm]{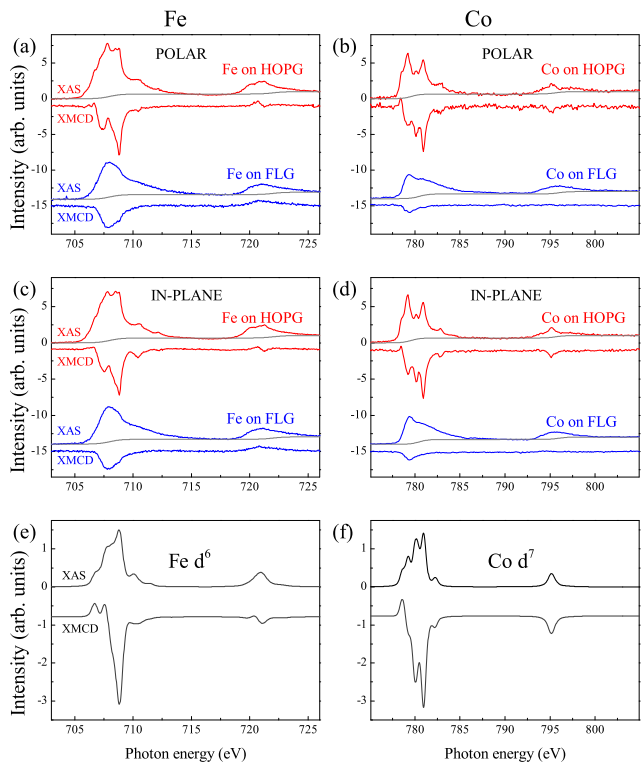}
\end{center}
\caption{\label{HOPG} XAS and XMCD spectra of Fe and Co adatoms after substrate background subtraction measured at $T=8$K and \textit{B}=5T for polar (a-b) and in-plane (c-d) geometries. Respective coverages are 0.020ML Fe and 0.006ML Co on HOPG, as well as 0.007ML Fe and 0.008ML Co on FLG. (e-f) show simulated XAS and XMCD spectra for pure $d^6$ and $d^7$ free atomic configurations.}
\end{figure}

Here, we use x-ray magnetic circular dichroism (XMCD) and x-ray absorption spectroscopy (XAS) techniques to determine the \textit{d}-shell occupation and magnetism of Fe, Co, Ni and Cu on highly oriented pyrolitic graphite (HOPG) and few-layer graphene grown on the C-terminated face of SiC [SiC(000-1)] (FLG). FLG with rotationally disordered stacking of more than five carbon layers provides a neutral surface with a Dirac dispersion, which is the closest to ideal graphene~\cite{Varchon}. \newline
We find that on HOPG, Fe and Co retain their free-atom electronic configuration corresponding to $d^n$ high-spin states with $n=6$ and $n=7$, respectively. On FLG, however, Fe and Co prefer a $d^{n+1}$ configuration, which is clearly reflected in the different x-ray spectral shapes compared to the free-atom counterparts. The strong differences found for the two multi-layer graphene substrates suggests that long-range interactions affect the adsorption of TMs and determine their ground state. For both graphene and HOPG we identify the bonding mechanism as physisorption, since no charge transfer between the guest TM and the host carbon atoms occurs.\newline
For the later 3\textit{d} elements Ni and Cu, instead, no substantial differences are found on graphene and HOPG. Surprisingly, Cu shows pronounced whiteline intensities, concomitant with a non-filled 3$d$ shell, but little magnetic susceptibility.  This behavior is attributed to the joint action of  $d$-$d$ intra-atomic Coulomb repulsion and charge fluctuation at guest atom sites. According to the electronic configuration, we describe the bonding as chemisorption for Cu and a mixed phase of chemisorbed and physisorbed species for Ni.\newline
%
\section{Methods}
Measurements were performed at the beamline ID08 of the European Synchrotron Radiation Facility using circularly polarized light tuned to the 3$d$ element $L_{2,3}$ absorption edges.\newline
TMs were deposited \textit{in-situ} onto FLG and HOPG at $T=8$K inside the magnet chamber by e-beam evaporators. TM coverages in units of monolayers (MLs) are given with respect to their bulk close-packed surfaces.\newline
Magnetic fields of $B=5$T were applied along the x-ray beam, both making an angle $\Theta$ relative to the surface normal. The XAS (XMCD) signal is defined as the average (difference) between positive and negative circularly polarized absorption spectra, which were measured at $T=8$K in the total electron yield mode for polar ($\Theta=0^{\circ}$) and in-plane ($\Theta=70^{\circ}$) geometries. All spectra shown in this work correspond to a maximum total x-ray exposure time of 2 minutes, which minimizes observed beam-induced time effects.
For details of sample preparation and measuring procedure we also refer to the supplementary information (SI).\newline
XAS and XMCD spectra are simulated using charge transfer multiplet theory~\cite{deGrootCT}. The numerical implementation is described in Ref.~\cite{Stepanow2}.
In our simulations we model the effect of electron charges at the surrounding carbon sites of the honeycomb network of graphene by a trigonal crystal field (CF) with symmetry $C_{6v}$ for hollow sites and $C_{3v}$ for top sites (see SI for a definition of the relevant CF parameters).
%
%
\section{Results}
Fig.~\ref{HOPG} shows experimental XAS and XMCD spectra of Fe and Co impurities on HOPG and FLG in polar (a-b) and in-plane (c-d) geometries. A rich multiplet structure is observed on HOPG as in the case of single atoms weakly interacting with the substrate~\cite{Gambardella alkali}. Multiplets in XAS originate from the core-hole interaction with the valence shell during photo-electron excitation, which serves as a fingerprint of the electronic configuration of single atoms~\cite{deGroot}. Fe and Co spectra on HOPG strongly resemble the simulated spectra for free $d^6$ and $d^7$ atoms displayed in Fig.~\ref{HOPG}(e-f). Equal XAS spectral shapes and intensities measured in polar and in-plane geometries further support the interpretation of free atom-like electronic configurations.\newline
On FLG instead, the spectra of Fe and Co atoms are strikingly different: The multiplet structure is less pronounced and the intensity of the XMCD with respect to XAS has decreased while the overall XAS and XMCD lineshapes remain nearly isotropic. These effects arise from a screening of the above mentioned core-hole interaction as well as a decreased magnetic moment in the 3$d$ shell. This suggests an enhanced atom-substrate interaction and a modification of the electronic 3\textit{d} configuration since the spectra are not merely a result of smearing out the spectral intensity. These properties and similar spectral shapes are also found in Ref.~\cite{Elbo} and therefore seem to be a benchmark for Co and Fe on graphene on SiC substrates.\newline
%
\begin{figure}
\begin{center}
\includegraphics[trim=0 0 0 0,clip,width=6.9cm]{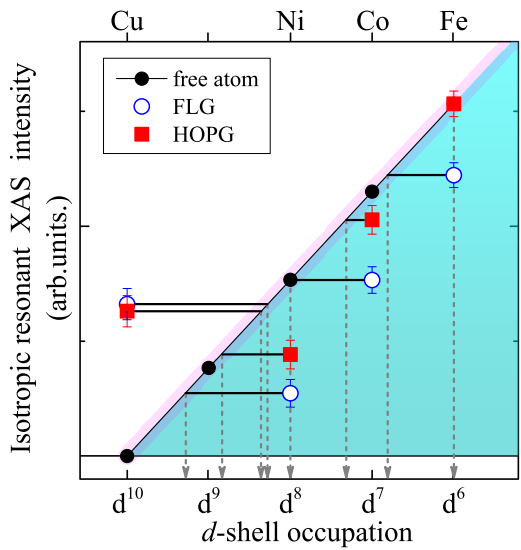}
\end{center}
\caption{\label{plot} Isotropic resonant XAS intensity of Fe, Co, Ni, and Cu adatoms plotted versus their Hartree-Fock free atom configurations $d^n$. The vertical dashed lines indicate the respective $d$-shell occupations $n_d$ on HOPG and FLG.}
\end{figure}
%
For a quantitative analysis, we use the first sum rule~\cite{Stöhr}, which directly correlates the integral resonant $L_{2,3}$ XAS intensity with the number of \textit{d} electrons $n_d$. The XAS intensity corresponding to resonant transitions into \textit{d} states is obtained upon subtraction of a step function from the measured XAS (grey lines underneath the experimental XAS data in Fig.~\ref{HOPG}), which takes into account transitions into continuum states. For a direct comparison of the different samples, we have normalized the edge jump of all spectra according to their element specific x-ray absorption cross section, which vary by up to 20$\%$~\cite{correction}. Using the pure $d^6$ Fe configuration found on HOPG as a reference, the number of holes of all elements can be estimated. In Fig.~\ref{plot}, the isotropic resonant XAS intensity~\cite{note1} is plotted versus the integer $d$-shell occupation of their respective free atom Hartree-Fock configuration. Data points that are above (below) the line indicate a decreased (increased) $n_d$ with respect to the free atom.
%
\subsection{Discussion: Fe and Co}
\begin{figure}
\begin{center}
\includegraphics[trim=13 0 0 0,clip,width=8.0cm]{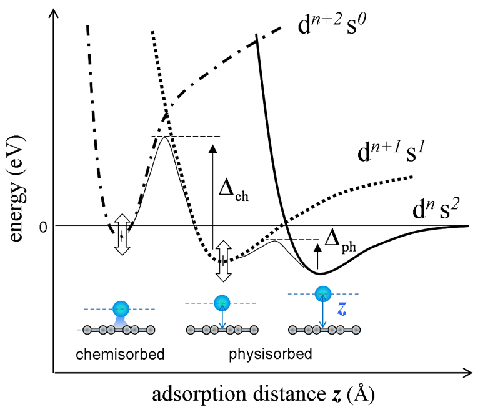}
\end{center}
\caption{\label{rudenko} Schematic potential landscape for physi- and chemisorption of Co and Fe adatoms on graphene. The relative energies of the potentials of the three adsorption configurations can shift with respect to each other for different elements and substrates (indicated by the vertical double arrows), which also affects the barriers $\Delta_{\text{ph}}$ (between physisorbed configurations) and the activation barrier $\Delta_{\text{ch}}$ towards chemisorption. All energies are given with respect to the $d^ns^2$ ground state in the free atom limit.}
\end{figure}

The pronounced variations in $n_d$ for Co and Fe on HOPG and FLG point towards a different bonding configuration on the two substrates, consistent with a recent many-body, quantum chemical study of Co on graphene~\cite{Rudenko}. This study predicts two equilibrium positions, at $z\sim3.1$\AA\ and $z\sim2.3$\AA\  corresponding to physisorbed high- and low-spin configurations with $d^n$ and $d^{n+1}$ occupations, plus a chemisorbed $d^{n+2}$ configuration at higher energies (see sketch in Fig.~\ref{rudenko}). We suggest that on HOPG, for both Co and Fe atoms, the activation barrier $\Delta_{\text{ph}}$ from high- to low-spin configurations is higher than on FLG, so that impinging Co and Fe atoms are trapped in the first potential minimum~\cite{note2}. On FLG instead, the $d^{n+1}$ configuration appears to be more favorable for both Fe and Co, resulting in $d^7$ and $d^8$ electronic configurations, respectively.

Our suggested substrate dependent activation barriers for TM physisorption raises the question on the detailed mechanism distinguishing the seemingly similar multi-layer graphene samples HOPG and FLG. We can exclude significant carbon band filling effects, since our results on neutral FLG are very similar to those in Ref.~[\onlinecite{Elbo}] for \textit{n}-doped graphene. Moreover, we can  rule out possible differences in the carbon $p_z$ electrons Coulomb screening, since according to Ref.~\cite{correlation} multi-layer graphene should already behave similar to graphite.\newline
Long-range forces are a better candidate, and they are also consistent with the larger bonding distances for the physisorbed species. Indeed, it was recently shown that the substrate underneath graphene can greatly influence the equilibrium position of adatoms or molecules e.g. via van der Waals interactions~\cite{NO2}. While in our case we do not expect strong differences in the van der Waals forces for TMs on HOPG and FLG (contributions due to the proximity of the SiC substrate should strongly decay through 5 graphene spacer layers), electrostatic effects due to charge defects and substrate morphology at the SiC interface are more plausible due to their larger decay length. Native charged defects at the interface between SiC(000-1) and the first graphene layer have been observed by STM in correspondence to the (2$\times$2) reconstruction~\cite{Fanny1, Fanny2}. Similarly, for graphene grown on the Si-terminated SiC substrates, the interface is known to be quite disordered~\cite{Crommie2}, leading to surface charges and low energy interface states~\cite{Mallet,c,d} which contribute to graphene doping and give rise to long range effects.

At this point, we want to mention that for the case of FLG, our experimental $d$-shell occupations for Fe and Co are in agreement with previous DFT+\textit{U} calculations on free-standing graphene~\cite{Wehlig}. These calculations also predict Fe and Co adatoms to favor top adsorption sites, which is consistent with the spectral shapes found on FLG. In fact, as evinced from the strong splitting and mixing of $|m|=2$ and $|m|=1$ states in the DOS of Co for top sites, the latter correspond to large cubic crystal fields along the $C_3$ symmetry axis of the octahedron~\cite{Wehlig}. According to our multiplet calculations, such values likely lead to isotropic spectral shape, as observed in our data (see SI). Conversely, the t$_{2g}$ level is hardly split in energy and the lower levels a$_1$ and $e_g$ are nearly degenerate. Therefore, for bridge and hollow adsorption sites, we would expect significant XAS line-shape anisotropies.
%
\subsection{Discussion: Ni and Cu}

\begin{figure}
\begin{center}
\includegraphics[trim=0 0 0 0,clip,width=8.5cm]{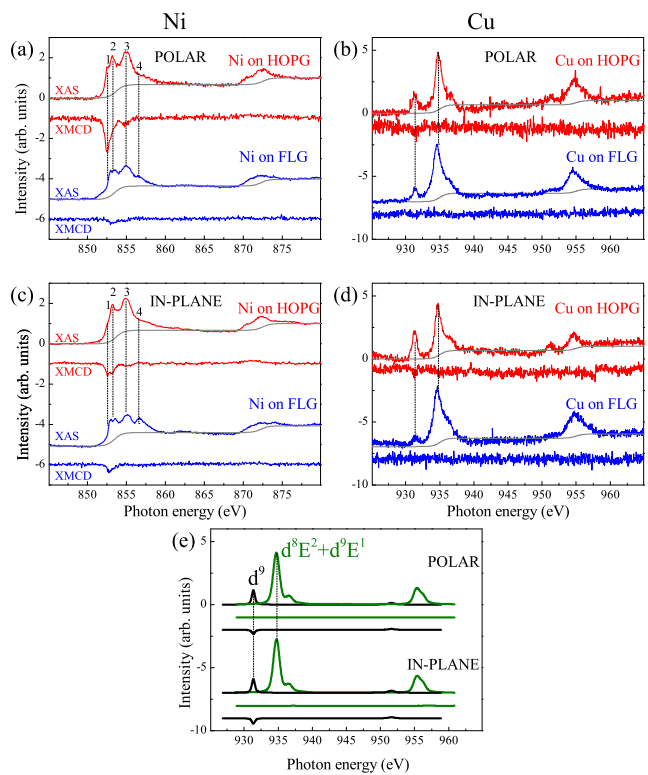}
\end{center}
\caption{\label{FLG} XAS and XMCD spectra of Ni and Cu adatoms after substrate background subtraction  measured at $T=8$K and \textit{B}=5T for polar (a-b) and in-plane (c-d) geometries. Respective coverages are 0.017ML for Ni and 0.005ML Cu atoms on HOPG, as well as 0.023ML Ni and 0.005ML Cu on FLG. Simulated XAS and XMCD spectra are shown in (e) for an incoherent superposition of Cu $d^9$ with crystal field (D$_{3q}=1$eV, D$_{\sigma}=-0.04$eV, D$_{\tau}=-0.02$eV) and Cu $d^8E^2+d^9E^1$ with crystal field (D$_{3q}=2$eV, D$_{\sigma}=0.14$eV, D$_{\tau}=0.08$eV).}
\end{figure}

As notified in the introduction, a more complex behavior is observed towards the end of the 3$d$ transition row. In Fig.~\ref{FLG}, XAS and XMCD spectra for Ni and Cu on HOPG and FLG are shown, for polar (a-b) and in-plane (c-d) geometries. The observed differences in the spectral shapes on graphene and HOPG gradually disappear for the later 3$d$ elements Ni and Cu. This is further underlined by the electronic configuration shown in Fig.~\ref{plot}, where for both substrates, Ni TMs are found close to $d^9$ ($n_{d}\sim8.8$ on HOPG and $n_{d}\sim9.3$ on graphene), and Cu TMs between $d^8$ and $d^9$ (about 8.4 on both substrates). The reduced dependence on the substrate already hints that long-range interactions are less important for Ni and Cu. On the other hand it also shows that the theoretically predicted anomalous broadening of TM impurity levels due to interaction with Dirac Fermions~\cite{Co_graphene_kondo_mag} does not lead to a different behavior on FLG compared to HOPG.
We also stress that, very intriguingly, neither Ni nor Cu appear in the $d^{10}$ configuration proposed by theory~\cite{Duffy-Blackman, Johll, Wehlig,Cu,Cu2}. A pure $d^{10}$ configuration would give a simple step function in XAS.\newline
%
%
Modeling the Ni and Cu electronic structure is more complex compared to Fe and Co. Energy differences between ground and excited states are small, leading e.g. to complex ground states in metallic Ni with superposition of $d^8$, $d^9$ and $d^{10}$ configurations~\cite{Nimetal}, or the coexistence of $s^1d^9$ and $s^2d^8$ configurations for high-temperature Ni atoms in the gas phase~\cite{MartinsPRL}.

We first discuss the more transparent case of Cu. On both HOPG and FLG, Cu shows a spectrum comprising two well separated resonant contributions: a minority one at $\sim 931$eV with dichroism, and the predominant non-magnetic one at larger photon energy. The minority component has a slight angular dependence of the XAS and XMCD, which we attribute to a final crystal field potential. Its spectral shape resembles that of a pure Cu $d{^9}$ electronic configuration as shown in the simulations of Fig.~\ref{FLG}(e). We want to comment here that the minority $d{^9}$ species shows a pronounced beam-induced time effect, which enhances its spectral weight upon longer exposure times (see SI). \newline
The predominant non-magnetic component is instead unusual. Here, the observation of a strong whiteline intensity indicates a significant hole occupation in the $d$-shell, however, the overall magnetic moment of the system is zero. As we are going to show, this finding can be understood only in the framework of a charge transfer (CT) model accounting for charge fluctuations in the initial and final states~\cite{deGrootCT,Sebastian,Stepanow2}.

We model the effect of electron charges at the surrounding carbon sites by a trigonal CF with symmetry $C_{6v}$ for graphene hollow sites and $C_{3v}$ for top sites. We consider the Cu atoms in a $d^{8}$ configuration with two additional electrons in a separate orbital $E$, which can hop both on and off the atom. The groundstate wave function is then a coherent superposition $d^8E^2+d^9E^1$, where the additional electrons in the orbital $E$ represent the available states from graphene and from the TM itself. This scheme has an even number of electrons per configuration generating a spin singlet ground state. The XAS simulation shown for Cu in Fig.~\ref{FLG}(e) (green curve) fit our data very well and lead to non-magnetic ground states for both hollow and top sites. The fit corresponds to a $d$-shell occupation of $n_d\sim8.1$, which is close to the independently derived value in Fig.~\ref{plot}.

We emphasize the fact that using a $d^8$ configuration with CF but without CT, cannot account for the observed magnetic quenching, and moreover does not reproduce the XAS spectral shape. A $d^9$ configuration would appear as a more likely scenario, however a finite XMCD would be found in any CF and neither the XAS spectral shape (see simulation in Fig.~\ref{FLG}(e)) nor the $d$-shell occupation would match the experimental results. Finally, if $sd$ hybridization~\cite{sd} between the symmetry related $4s$ and $3d_z^2$ orbital would occur, the 11 electrons of Cu ($s^1d^{10}$) would be re-distributed in 12 orbitals, leading to a certain hole occupation with $d$ character in XAS. However, also in this case the $d$-shell occupation would be larger than what is found experimentally and the XAS shape similar to $d^9$.

Since the 4$s$ states are expected to be high in energy in the vicinity of graphene and half-filled~\cite{Cu}, the orbitals $E$ forming the coherent superposition $d^8E^2+d^9E^1$ required in our model can only be the carbon $p_z$ states. This scenario coherently explains the large XAS $d$ resonance in Figs.~\ref{FLG}(b, d), the magnetic moment quenching, and the average hole occupation $n_d=8.4$ found in the experiments.\newline
The proposed electronic configuration reveals a tendency of Cu towards delocalization, that can be qualitatively explained as a means to relieve the strong Coulomb repulsion between \textit{d} electrons. Since this non-magnetic species with a large $d$ resonance presumes a static charge exchange with the substrate, the adatom-substrate interaction can be labeled as chemisorption.

Turning to the Ni adatom case, we note that the XAS spectral shapes shown in Fig.~\ref{FLG} are similar on HOPG and FLG, consistent with the comparable Ni $d$-shell occupancies $n_d$ in Fig.~\ref{plot}. On both substrates XMCD signals are observed, however, the one on FLG is strongly reduced with respect to HOPG. Such substrate dependent magnetic properties are reminiscent of the earlier elements Fe and Co in the 3$d$ row, however they are far less prominent in the case of Ni. The overall light substrate dependent magnetic properties of Ni adatoms and similarities to the Cu spectral shapes suggests an interpretation analogous to Cu.\newline
Similar to Cu, the Ni experimental XAS and XMCD spectra in Fig.~\ref{FLG}(a, c) appear to contain at least two independent species, leading to spectral weights with large (peaks 1-2) and small (peaks 3-4) magnetic dichroism. The sharp single peak (peak 1) in the XMCD at lowest photon energies, particularly pronounced on HOPG, resembles a $d^9$ component as shown in Fig.~\ref{FLG}(e) for the case of Cu. It could be interpreted as physisorbed Ni with a $d^9$ configuration but, as seen from the angular dependence, sensitive to the crystal field of graphene. Similar to the minority Cu $d^9$ species this Ni configuration shows an increased susceptibility to beam-induced effects, however here the spectral weight decreases upon exposure (see SI). On the other hand, analogous to the case of Cu, we ascribe the components (3-4) to a coherent superposition $d^8E^2+d^9E^1$ with an almost completely quenched magnetic moment.\newline
We want to comment that even though our experimental data resemble those in Ref.~[\onlinecite{Hamburg2}] for Ni on graphene on the Si-face of SiC, we arrive to a different interpretation employing the concept of charge fluctuation, which explains the combination of prominent Ni 3$d$ whiteline intensities and suppressed magnetic susceptibilities. The key to the conceptual understanding is the Cu case, which serves as the model system in the strongly hybridized limit. We also note that the existence of two Ni species with different magnetic behavior is consistent with the adsorption scheme proposed in Fig.~\ref{plot}.

Finally, we address the question to what extent DFT calculations can account for the formation of finite Ni moment, considering that
earlier DFT calculations predicted Ni adatoms on graphene to be non-magnetic~\cite{Duffy-Blackman, Johll, Wehlig,Elbo}. Our DFT calculations (see SI) suggest that shortcomings are connected with the inability of common DFT schemes to describe
many-body interactions which can govern the adsorption geometry. Namely, if DFT-derived adatom-graphene vertical distances
$z_{\text{DFT}}$ are increased by 25\%, finite Ni moments form, provided that individual adatoms indirectly feel their proximity
at intermediate lateral spacings in the range of 5\AA\ $< d_{\text{Ni-Ni}} < 10$\AA.
\section{Conclusions}
3$d$-shell electronic configurations of Fe, Co, Ni and Cu adatoms on multi-layer graphene systems are determined for the first time directly from experimental x-ray absorption spectroscopy data. The combination of experiment and simulations shine light on the long-debated nature of substrate dependent interactions between transition metals and supported graphene surfaces.\newline
On highly-ordered pyrolitic graphite Fe and Co are found in weakly bound high-spin configurations as predicted by quantum chemistry calculations including long-range interactions. Density functional theory calculations so far do not predict such findings, which underlines shortcomings regarding the simulation of adsorption processes on realistic graphene systems used in experiments.\newline
For the late 3$d$ elements Ni and Cu we show that the ground state can be well described by a coherent mixture of $d^8$ and $d^9$ electronic configurations, which are subject to charge fluctuations. The complex case of Ni and Cu represents further motivation for \textit{ab-initio} many-body calculations.\newline

We acknowledge ESRF for allocating beamtime and GACR for support (project 108/11/0853). We thank A. Fondacaro for technical help; J. Vackar and A. Simunek for assistance with the ABINIT code; A. R. acknowledges support from EU Graphene Flagship (contract No 604391).


\begin{thebibliography}{}
\bibitem{spin1} N. Tombros, C. Jozsa, M. Popinciu, H. T. Jonkman, and B. J. van Wees, Nature (London) \textbf{448}, 571 (2007).
\bibitem{mobility}X. Du, I. Skachko, A. Barker, and E. Y. Andrei, Nature Nanotechnology \textbf{3}, 491 (2008).
\bibitem{spin2}  W. Han, K. Pi, K. M. McCreary, Y. Li, J. J. I. Wong, A. G. Swartz, and R. K. Kawakami, Phys. Rev. Lett. \textbf{105}, 167202 (2010).
\bibitem{spin injection} J. Maassen, W. Ji, and H. Guo, Nano Lett. \textbf{11}, 151 (2011).
\bibitem{magnetoresistance} E. Cobas, A. L. Friedman, O. M. J. van't Erve, J. T. Robinson, and B. T. Jonker, Nano Lett., \textbf{12}, 3000 (2012).
\bibitem{Castro-Neto} A. H. Castro Neto, F. Guinea, N. M. R. Peres, K. S. Novoselov, and A. K. Geim, Rev. Mod. Phys. \textbf{81}, 109-162 (2009).
\bibitem{Sengupta} K. Sengupta and G. Baskaran, Phys. Rev. B \textbf{77}, 045417 (2008).
\bibitem{Hamburg1} M. Gyamfi, T. Eelbo, M. Wasniowska, and R. Wiesendanger, Phys. Rev. B \textbf{84} 113403 (2011).
\bibitem{Crommie} V. W. Brar, R. Decker, H.-M. S., Y. Wang, L. Maserati, K. T. Chan, H. Lee, C. O. Girit, A. Zettl, S. G. Louie, M. L. Cohen, and M. F. Crommie, Nature Physics \textbf{7}, 43 (2011).
\bibitem{Elbo} T. Eelbo, M. Wasniowska, P. Thakur, M. Gyamfi, B. Sachs, T. O. Wehling, S. Forti, U. Starke, C. Tieg, A. I. Lichtenstein, and R. Wiesendanger, Phys. Rev. Lett. \textbf{110}, 136804 (2013).
\bibitem{graphene} T. Eelbo, M. Wasniowska, M. Gyamfi, S. Forti, U. Starke, and R. Wiesendanger, Phys. Rev. B \textbf{87}, 205443 (2013).
\bibitem{Hamburg2} M. Gyamfi, T. Eelbo, M. Wasniowska, T. O. Wehling, S. Forti, U. Starke, A. I. Lichtenstein, M. I. Katsnelson, and R. Wiesendanger, Phys. Rev. B \textbf{85}, 161406(R) (2012).
\bibitem{Co_HOPG} P. K. J. Wong, M. P. de Jong, L. Leonardus, M. H. Siekman, and W. G. van der Wiel, Phys. Rev. B \textbf{84}, 054420 (2011).
\bibitem{Brune}F. Donati, Q. Dubout, G. Autes, F. Patthey, F. Calleja, P. Gambardella, O.V. Yazyev, and H. Brune, Phys. Rev. Lett.~\textbf{111}, 236801 (2013).
\bibitem{a} J. Klimes and A. Michaelides, J. Chem. Phys. \textbf{137}, 120901 (2012).
\bibitem{b} A. N. Rudenko, F. J. Keil, M. I. Katsnelson, and A. I. Lichtenstein  Phys. Rev. B \textbf{84}, 085438 (2011); A. N. Rudenko, F. J. Keil, M. I. Katsnelson, and A. I. Lichtenstein, Phys. Rev. B \textbf{83}, 045409 (2011).
\bibitem{Xiao} R. Xiao, D. Fritsch, M. D. KuzÕmin, K. Koepernik, and M. Richter, Phys. Rev. B \textbf{82}, 205125 (2010).
\bibitem{Duffy-Blackman} D. M. Duffy and J. A. Blackman, Phys. Rev. B \textbf{58} 7443(1998).
\bibitem{Johll} H. Johll, H. C. Kang, and E. S. Tok, Phys. Rev. B \textbf{79}, 245416 (2009).
\bibitem{Krasheninnikov} A. V. Krasheninnikov, P. O. Lehtinen, A. S. Foster, P. Pyykkö, and R. M. Nieminen, Phys. Rev. Lett. \textbf{102}, 126807 (2009).
\bibitem{Wehlig} T. O. Wehling, A. I. Lichtenstein, and M. I. Katsnelson, Phys. Rev. B \textbf{84}, 235110 (2011).
\bibitem{Rudenko} A. N. Rudenko, F. J. Keil, M. I. Katsnelson, and A. I. Lichtenstein, Phys. Rev. B \textbf{86}, 075422 (2012).
\bibitem{Varchon} F. Varchon, P. Mallet, L. Magaud, and J.-Y. Veuillen, Phys. Rev. B \textbf{77}, 165415 (2008).
\bibitem{deGrootCT} F. de Groot, Coordination Chemistry Reviews \textbf{249}, 31 (2005).
\bibitem{Stepanow2} S. Stepanow, A. Mugarza, G. Ceballos, P. Moras, J. Cezar, C. Carbone, and P. Gambardella,
Physical Review B, \textbf{82}, 014405 (2010).
\bibitem{Gambardella alkali} P. Gambardella, S. S. Dhesi, S. Gardonio, C. Grazioli, P. Ohresser, and C. Carbone, Phys. Rev. Lett. \textbf{88}, 047202 (2002).
\bibitem{deGroot} F. M. F. de Groot, J. C. Fuggle B. T. Thole, G. A. Sawatzky, Phys. Rev. B. \textbf{42}, 5459 (1990).
\bibitem{Stöhr}J. St\"ohr, and H. K\"onig, Phys. Rev. Lett. \textbf{75}, 3748 (1995).
\bibitem{correction} B. L. Henke, E. M. Gullikson, and J. C. Davis, At. Data Nucl. Data Tables \textbf{54}, 181 (1993).
\bibitem{note1}The isotropic intensity $I_0$ has been calculated as $I_0=I_z+2I_P$ where $I_z$ and $I_P$ are the grazing ($\Theta=90^\circ$) and polar XAS intensities. The former can be calculated from the XAS intensities measured polar and in-plane.
\bibitem{note2}The slight deviations from the simulated spectra on HOPG at low photon energies are probably due to a statistical amount of minority species with $d^{n+1}$ and $d^{n+2}$ configurations.
\bibitem{correlation}T. O. Wehling, E. \ifmmode \mbox{\c{S}}\else \c{S}\fi{}a\ifmmode \mbox{\c{s}}\else \c{s}\fi{}\ifmmode \imath \else \i \fi{}o\ifmmode \breve{g}\else \u{g}\fi{}lu, C. Friedrich, A. I. Lichtenstein, M. I. Katsnelson, and S. Bl\"ugel, Phys. Rev. Lett. \textbf{106}, 236805 (2011).
\bibitem{NO2} J. Dai and J. Yuan, Chemical Physics \textbf{405}, 61 (2012).
\bibitem{Fanny1}F. Hiebel, P. Mallet, F. Varchon, L. Magaud, and J-Y. Veuillen, Phys. Rev. B \textbf{78}, 153412 (2008).
\bibitem{Fanny2} F. Hiebel, P. Mallet, J.-Y. Veuillen, and L. Magaud, Phys. Rev. B \textbf{83}, 075438 (2011).
\bibitem{Crommie2}  V. W. Brar, Y. Zhang, Y. Yayon, T. Ohta, J. L. McChesney, A. Bostwick, E. Rotenberg, K. Horn, and M. F. Crommie, Appl. Phys. Lett. \textbf{91}, 122102 (2007).
\bibitem{Mallet} P. Mallet, F. Varchon, C. Naud, L. Magaud, C. Berger, and J.-Y. Veuillen, Phys. Rev. B \textbf{76}, 041403(R) (2007).
\bibitem{c} C. Coletti, C. Riedl, D. S. Lee, B. Krauss, L. Patthey, K. von Klitzing, J. H. Smet, and U. Starke, Phys. Rev. B. \textbf{81}, 235401 (2010).
\bibitem{d} C. Riedl, C. Coletti and U. Starke, J. Phys. D: Appl. Phys. \textbf{43}, 374009 (2010).
\bibitem{Co_graphene_kondo_mag}B. Uchoa, V. N. Kotov, N. M. R. Peres, and A. H. Castro Neto, Phys. Rev. Lett. \textbf{101}, 026805 (2008).
\bibitem{Cu} M. Wu, En-Zuo Liu, M. Y. Ge, J. Z. Jiang, App. Phys. Lett. \textbf{94}, 102505 (2009).
\bibitem{Cu2}M. Amft, S. Lebegue, O. Eriksson, and N. V. Skorodumova, J. Phys.: Condens. Matter \textbf{23}, 395001 (2011).
\bibitem{Nimetal}T. Jo and G. A. Sawatzky, Phys. Rev. B \textbf{43}, 8771 (1991).
\bibitem{MartinsPRL} K. Godehusen,T. Richter, P. Zimmermann, and M. Martins, Phys. Rev. Lett. \textbf{88}, 217601 (2002).
\bibitem{Sebastian} S. Stepanow, P. S. Miedema, A. Mugarza, G. Ceballos, P. Moras, J. C. Cezar, C. Carbone, F. M. F. de Groot, and P. Gambardella, Phys. Rev. B \textbf{83}, 220401(R) (2011).
\bibitem{sd} Modern Electronic Structure Theory, David R. Yarkony, Ed. World Scientific Publishing, Singapore (1995).

\end{thebibliography}
\end{document}